# Optical Design of the Orbiting Astronomical Satellite for Investigating Stellar Systems (OASIS)

**Siddhartha Sirsi*[a,b], Yuzuru Takashima[b], Arthur Palisoc[c], Heejoo Choi[b,d], Jonathan W. Arenberg[e], Daewook Kim[a,b,d] , Christopher Walker[a,b]**

[a]Department of Astronomy and Steward Observatory, University of Arizona, 933 N. Cherry Ave., Tucson, AZ 85721, USA
[b]Wyant College of Optical Sciences, University of Arizona, 1630 E. University Blvd., Tucson, AZ 85721, USA
[c] L'Garde, Inc., 15181 Woodlawn Avenue, Tustin, CA 92780, USA
[d]Large Binocular Telescope Observatory, University of Arizona 933 N Cherry Avenue, Tucson, AZ 85721, USA
[e] Northrop Grumman Systems Corporation, 1 Space Park Blvd, Redondo Beach, CA 90278

**Abstract**. The Orbiting Astronomical Satellite for Investigating Stellar Systems (OASIS) is a proposed space telescope with a 14 m inflatable primary reflector that will perform high spectral resolution observations at terahertz frequencies with heterodyne receivers. The telescope consists of an inflatable metallized polymer membrane that serves as the primary antenna, followed by aberration correction optics, and a scanner that enables a 0.1 degrees Field of Regards while achieving diffraction limited performance over wavelength range from 63 to 660 μm.

Here the parametric solution space of the OASIS inflatable telescope design is systematically investigated by establishing analytical relations among figure of merits including 1$^{st}$ order geometrical photon collection area and the size of correction optics. The 1$^{st}$ order solution was further optimized by ray-trace code by incorporating numerically calculated mirror shape with pre-formed membrane gores. Design study shows that a space-based telescope with an effective photon collection area of over 90m$^2$ can be achieved utilizing a 14m inflatable aperture.

## 1 Introduction

Water is an essential ingredient to the origin and evolution of life on Earth[1]. Water also plays an important role in the formation of planets. The Orbiting Astronomical Satellite for Investigating Stellar Systems (OASIS) is a proposed space-based telescope with a 14 m diameter inflatable primary reflector/antenna that will follow the water trail from galaxies to oceans by performing high spectral resolution observations of water at terahertz frequencies[2,3]. The telescope's primary reflector consists of transparent and metallized polymer membranes sealed around their periphery, constrained by a tensioning structure, and inflated to the required pressure (Fig. 1). The large inflatable primary antenna (A1) is the key to achieving the large collecting areas required for the proposed observational study. OASIS will have ~10× the collecting area and 6× the angular resolution of the Herschel Space Observatory[4] and compliments the short wavelength capabilities of James Webb Space Telescope (0.6 to 28.3 μm). Such a large aperture is realized by utilizing a lightweight, stowable polymer-based primary antenna that reduces launch cost, as well as the lead time required for fabrication.

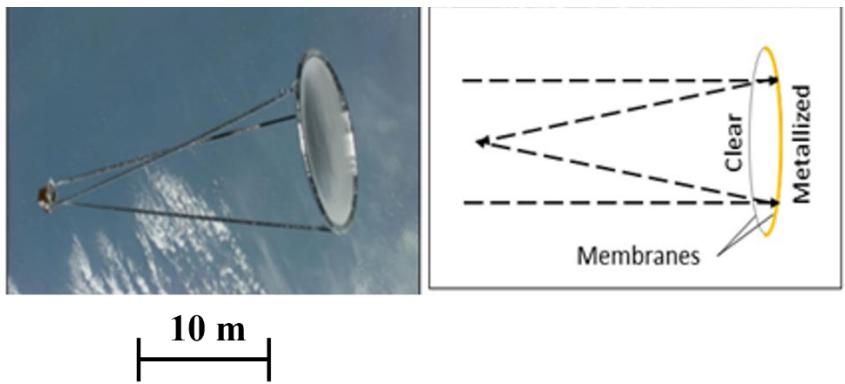

**Fig. 1** Left: Inflatable Aperture Experiment (IAE) demonstrated a 14 m inflatable aperture in space (1996) for use at X-band. Right: Incoming signal focused by concave metallized membrane. [Image: NASA]

The shape of the primary antenna A1 is a function of pressure, material properties, and boundary conditions. The dependence of the shape of A1 on multiple parameters offers unique opportunities in achieving design goals. The shape of inflatable membranes has been discussed in several articles. A 4$^{th}$ order solution for uniformly loaded inflated monolithic membrane is reported by Hencky[5]. Surface shape with higher order terms is presented by Fichter[6]. Besides these analytic solutions, the Finite Element Analyzer for Membranes (FAIM) software package was developed by L'Garde Inc to numerically calculate the shape of A1 after inflation. This numerical approach was adopted to simulate the final inflated shape of A1 under different conditions[7].

Inflated membrane reflectors formed from flat dielectric sheets inherently form Hencky surfaces. Hencky surfaces are neither spherical or parabolic and involve coupled 2$^{nd}$ and 4$^{th}$ order terms. In this situation, the magnitude of the spherical aberration in A1 is coupled to its focal length, F/#. An approach for correcting spherical aberration and focus utilizing four mirrors is reported by Burge et al[8]. The OASIS' aberration correction mirror pair is designed to simultaneously tackle the on-axis spherical aberrations and off-axis aberrations, *i.e.,* coma, encountered with an inflatable. Moreover, the effective collection area of the telescope and secondary mirror sizes are also a function of the shape and size of A1. In addition, the wide wavelength range of OASIS requires appropriate tolerance budgeting.

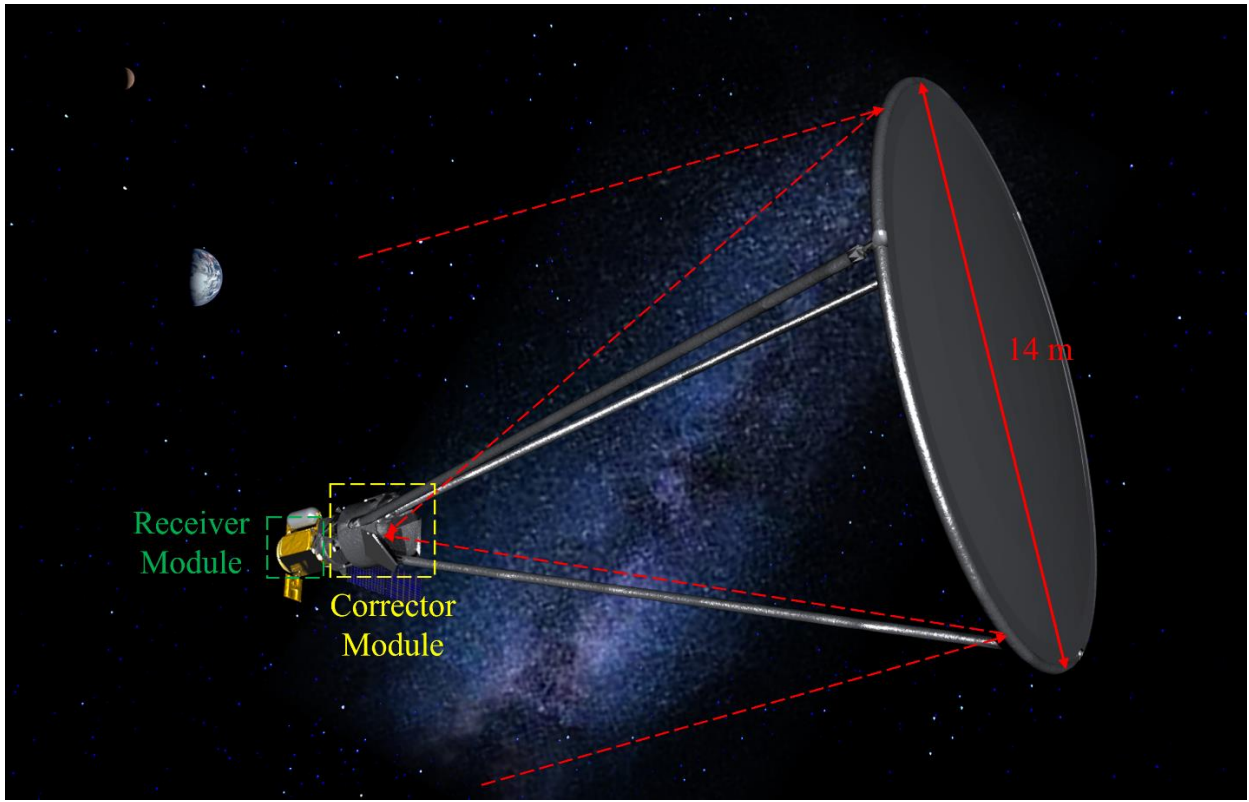

**Fig. 2** OASIS mission concept showing the corrector and receiver modules (left side of the figure) and the fully deployed 14 m diameter primary reflective antenna A1 (right side of the figure), which is an inflatable membrane optic.

Within the complex and mutually coupled design landscape, we have developed a $1^{st}$ order analytical model for inflatable reflectors. Based on the $1^{st}$ order model, further optimization of corrector optics, as well as tolerance analysis is performed as a function of collecting area and effective focal length (F#). This approach determines the optimal design that delivers the required effective collection area utilizing a compact aberration correcting mirror pair. The results are presented in the form of solution space contour plots, which serve as a powerful tool for converting scientific requirements to optical system specifications.

Here we report the design process of a large aperture telescope with an inflatable primary reflector. In Section 2 the OASIS optical specification is discussed in conjunction with the concept of operation. In Section 3, power arrangement, size requirement for correction optics, and photon collection area is addressed by developing an analytical model based on a $4^{th}$ order Hencky model. Further optimization of the corrector optics, incorporation of a numerical solution for surface shape, and a performance figure of merit is discussed in Section 4. Section 5 addresses photon collection area as a function of effective focal length (EFL) and F/# of the primary mirror in the design space while incorporating additional factors influencing on photon collection area, e.g., as-built Strehl intensity ratio and optical transmission. Section 6 discusses the optical design of a field of view (FoV) scanning mechanism for OASIS. The baseline design and optical path loss budget of a 14 m OASIS space telescope is discussed in Section 7, followed by a discussion of how to address challenges encountered in the design of inflatable optical systems.

## 2   OASIS Optical Specifications

The OASIS space telescope concept is illustrated in Fig. 2. Its 14 m primary reflector, A1, is initially stowed in the spacecraft and deployed in orbit using three expanding booms[9]. A1 is made up of two thin (~12 μm) polymer (e.g., Mylar or Kapton) membranes; one forming a clear canopy and the second an aluminized reflector. The space between the two membranes is pressurized to form the required concave reflective surface.

OASIS targets far-infrared (far-IR) transitions of water and its isotopologues, as well as HD and other molecular species, from 0.45 to 4.7 THz (660 to 63 μm) that are obscured by the Earth's atmosphere. The sensitivity required to detect and spectrally resolve these lines is provided by a large aperture coupled with state-of-the-art heterodyne receivers[10]. The terahertz heterodyne receivers need to be periodically (~20 s) chopped on and off targets. This translates to a minimum required field of view of 0.01 deg (6 arcmin). An F/16 system is selected to efficiently couple the telescope beam to the focal plane instruments. The key requirements concerning the optical design of OASIS are listed in Table 1.

**Table 1** Key optical design requirements of OASIS based on science goals and system architecture.

|  | Requirement |
| --- | --- |
| **F/#** | 16 |
| **Collection area** | >56 m$^2$ |
| **Field of view** | ±3 arcmin (circle) |
| **Payload Mass/Collection area** | 13 kg/m$^2$ |
| **Wavelength** | 63 – 660 μm |

## 3 Analytical Model of A1 with spherical aberration corrector

*3.1 Shape of inflatable primary antenna A1*

An inflatable mirror formed by pressurizing two thin, circular, monolithic flat polymer membranes bonded at the edges has a surface profile of an oblate spheroid whose figure can be expressed by an even power series known as Hencky Curve[11], given by

$$z(u) = \frac{D}{64F^2}\left(u^2 + 0.1111u^4\right), \tag{1}$$

where $D$ is the diameter of the mirror, $F$ is the $F/\#$, and $u = \dfrac{r}{D}$ is fractional radius. A Hencky surface will be assumed for A1 in the first order optical design.

Fig. 3 plots $z(u)$ for a Hencky surface with D = 20 m, and F = 1.25. For the purpose of comparison, parabolic sag for the same D and F/# is also shown. Because the sag value at the edge of mirror substantially deviates from the sag of parabola, a large amount of spherical aberration is induced by the Hencky surface.

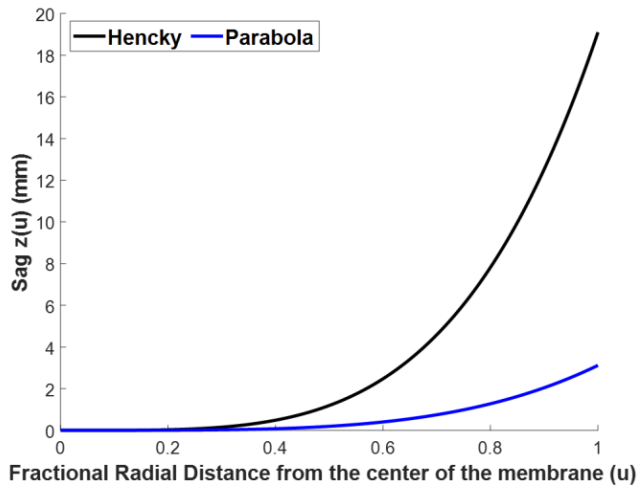

**Fig. 3** Comparison between Hencky and parabolic surface radial profile with the best fit sphere removed.

*3.2 First Order OASIS Optical Design Layout*

To correct for the spherical aberration due to the deviation of A1 from an ideal parabola, two concave mirrors, M2 and M3, are incorporated in the optical design as shown in Fig. 4. This mirror pair resides in the Corrector Module (Fig. 2). Together they corrects for both spherical aberration and off-axis aberrations (e.g., coma). The driving principles behind the design of the M2-M3 mirror pair are to

i. achieve the smallest possible mirror size,

ii. minimize the distance between mirror pairs while also minimizing the central hole diameter of the M2 and M3 mirrors, and

iii. maximize the geometrical photon collection area.

To satisfy those requirements, we adopted a power arrangement of the A1-M2-M3 system which is similar to that of a reflective null collector[12]. The M2 mirror is placed at the paraxial focus of A1 and corrects for the spherical aberration induced by A1. M3 relays paraxial focus of A1 back to M2, which is also the system focal point of A1-M2-M3. In this sense, M3 is a 1:1 relay optics that re-images the intermediate image formed by A1 to the location of M2 again while jointly correcting spherical aberration with M2. To maximize the effective collection area, the diameter of the central hole of M2 and M3 mirrors are minimized by placing M3 where the size of ray bundle of A1 is minimized or at the minimum blur position[13].

As depicted in Fig. 5, diameter ($D_1$) and focal length of A1 (or distance between A1 and M2: $d_{1-2}$) are chosen as primary design variables, since those determine the overall dimensions of the A1-M2-M3 optical system, which is constrained by science/mission requirements for aperture size and stow volume. Given $D_1$, the surface profile of A1 and $d_{1-2}$ is a function of pressure and membrane properties. For the parameters $D_1$ and $d_{1-2}$, viable M2-M3 corrector designs can be identified by using the approach described below.

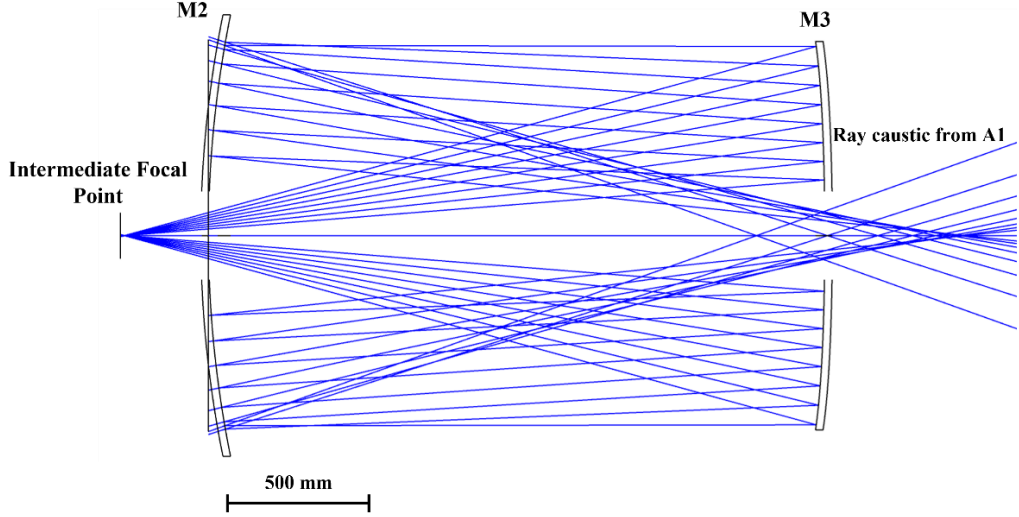

**Fig. 4** Optical layout of OASIS corrector M2-M3 mirror pair. Signal from the astronomical target reflects off of A1, passes through M3 hole, reflects off of M2, and is re-imaged by M3 through the M2 center hole.

*3.3 Design Space Survey Using Iterative Analytical Model*

The central obscuration of A1 ($h_{in}$) is dependent of M2, the M3 location, and the M3 entrance hole size. In turn, the location of M2 is dependent on the location of M3. An iterative analytical model was developed to optimize the first order power arrangement and solve for $h_{in}$, with its mutually coupled dependencies. This model determines the location and size of M2-M3 corrector optics, and the geometrical collection area of the telescope system by evaluating the value of $h_{in}$ for different combinations of A1 surface profiles and apertures. The 1$^{st}$ order design process is divided into sub steps 1 through 7 as follows.

**Step 1**: The Hencky surface profile is obtained for a given $D_1$, $d_{1-2}$, and field of view ($\theta$) with M2 hole diameter ($H_{M2}$).

$$H_{M2} = 2\tan(\theta)d_{1-2} + \frac{d_{IF}}{(f/\#)_{sys}}, \qquad (2)$$

where $(f/\#)_{sys}$ is system F/#, $d_{IF}$ is a distance from M2 mirror surface to focal plane of the A1-M2-M3 system, $d_{1-2} = \frac{R_1}{2}$, and $R_1$ is the base radius of curvature of A1. In Eq. (2), $d_{IF}/(f/\#)_{sys}$ is the ray bundle size at the surface of M2. The M2 hole size $H_{M2}$ accommodates the ray bundle footprint, while taking into account the shift of the ray bundle by an amount $2\tan(\theta)\, d_{1-2}$.

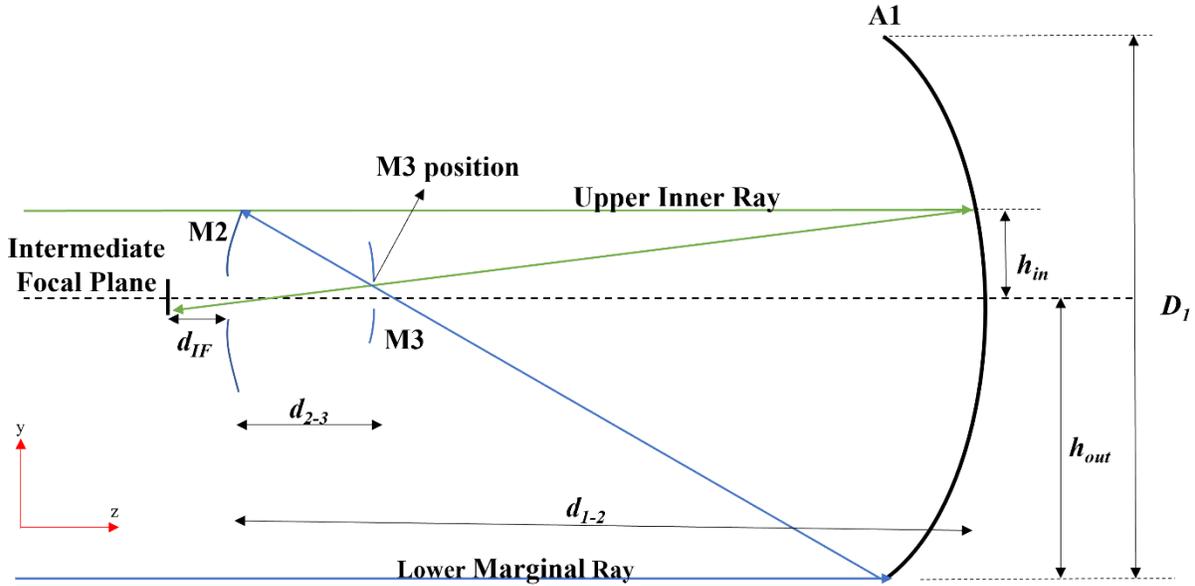

**Fig. 5** Analytical model showing the ray trace used to determine the systematic design space set by the mirror positions and critical dimensions of the corrector optics pair (M2 and M3).

**Step 2:** Calculate the height ($h_{in}$) of the incident ray (labeled as Upper Inner Ray in Fig. 5) at A1 which results in an image height of $H_{M2}/2$ at the initial intermediate focal plane. $h_{out} = D_1/2$ is the height of the marginal ray at A1 (labeled as Lower Marginal Ray in Fig. 5).

**Step 3:** The intersection of the Upper Inner Ray and the Lower Marginal Ray determines the location of M3 where the ray bundle size is minimized with spherical aberration of A1, as well as its hole size ($H_{M3}$).

**Step 4:** Given the distance $d_{2-3}$ between M2 and M3, the location of M2 is updated.

**Step 5:** The height of the lower marginal ray at the updated location of M2 determines the diameter of M2 ($D_{M2}$).

**Step 6:** Diameter of M3 ($D_{M3}$) is calculated by equating the $(f/\#)_{sys}$ with $(f/\#)_{M3}$ as

$$(f/\#)_{sys} = \frac{(R_1/2)}{2h_{out}}, \tag{3}$$

$$(f/\#)_{M3} = \frac{d_{2-3}+d_{IF}}{D_{M3}}, \tag{4}$$

$$D_{M3} = (d_{2-3}+d_{IF})\frac{2h_{out}}{R_1/2}, \tag{5}$$

**Step 7:** Geometrical photon collection area ($CA_{Geo}$) of the system is calculated.

$$CA_{Geo} = (h_{out}^2 - h_{in}^2)\pi. \tag{6}$$

The above steps are iterated until $h_{in}$ value converges.

An iterative analytical model is run using the parameters in Table 2.

Table 2 Iterative analytical model parameters.

| Model Parameters | Value |
| --- | --- |
| $R_1$ | 50 m |
| $D_1$ | [12 m 13 m 14 m 15 m 16 m 17 m] |
| $d_{2-3}$ | 2.2 m |
| $d_{IF}$ | 200 mm |
| $\theta$ | ±0.05° |

The results of the parametric sweep of $D_1$ for $R_1 = 50$ m and its comparison with the results calculated by geometrical raytracing software (ZEMAX, CodeV) are listed in Table 3. The shape of the M2 and M3 mirrors are optimized to provide diffraction limited performance.

The geometrical photon collection area of A1 calculated by the analytical model is in close accordance with the ZEMAX model. Table 3 shows that the iterative analytical model can be used to accurately predict the optical design parameters without having to go through the entire design process using ray tracing software. One important finding from the first order analysis is that Hencky surface profile of A1 requires M2 and M3 diameters exceeding 1 m.

**Table 3** Comparison of parametric sweep of D1 for R1 = 50 m between the analytical and ray tracing software (ZEMAX models).

| $D_1$ (m) | $d_{1\text{-}2}$ (m) | $D_{M2}$ (m) | $H_{M2}$ (mm) | $D_{M3}$ (m) | $H_{M3}$ (mm) | Analytical Geometrical Photon Collection Area (m$^2$) | Geometrical Photon Collection Area by ZEMAX (m$^2$) |
|---|---|---|---|---|---|---|---|
| 12 | 25.7705 | 1.33 | 141 | 1.15 | 134 | 102.55 | 100.31 |
| 13 | 25.5295 | 1.52 | 149 | 1.24 | 208 | 116.62 | 116.3 |
| 14 | 25.2675 | 1.74 | 156 | 1.34 | 301 | 129.62 | 129.21 |
| 15 | 24.9872 | 1.97 | 164 | 1.44 | 410 | 141.49 | 140.91 |
| 16 | 24.6881 | 2.23 | 171 | 1.53 | 533 | 152.29 | 151.72 |
| 17 | 24.3737 | 2.51 | 179 | 1.63 | 665 | 162.61 | 161.67 |

*Note)* $D_1$: Entrance pupil diameter of A1, $d_{1\text{-}2}$: Distance between A1 and M2, $D_{M2}$: Diameter of M2, $H_{M2}$: Hole diameter of M2, $D_{M3}$: Diameter of M3, and $H_{M3}$: Hole Diameter of M3

## 4  Characterization of A1 Surface Profile Figure of Merit

An iterative analytical model can accurately predict the size of M2 and M3 mirrors, and the geometrical photon collection area. The model is applied to different combinations of A1 profiles and entrance pupil diameters to determine suitable A1 surface profiles which minimizes the M2-M3 mirror sizes while satisfying the science requirement for collecting area. An alternative to forming the primary mirror from a flat membrane is to utilize several pre-formed slices of membranes (gores) that are stitched together to create the desired surface profile. L'Garde Inc provided surface profile data for quasi-parabolic shaped reflectors formed from gores[7]. These were compared with the case of an ideal parabola and Hencky surface. The results are shown in Fig. 6

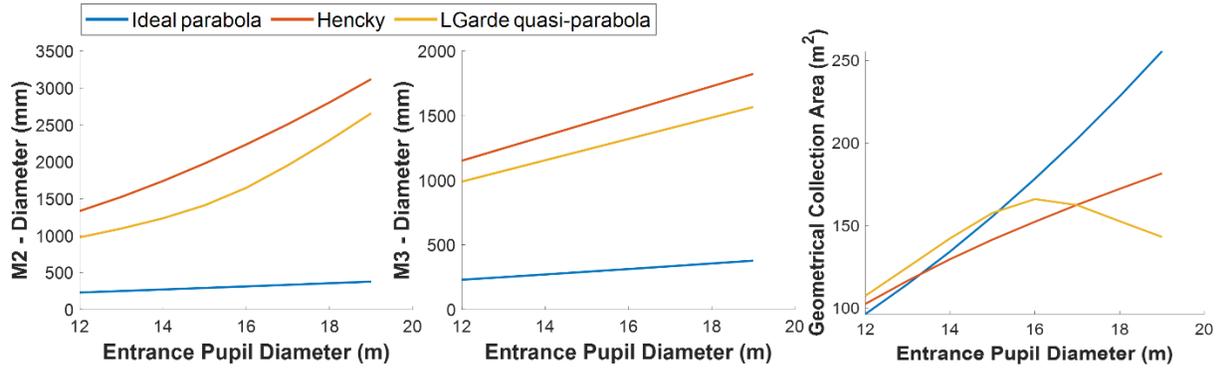

**Fig. 6** Comparison between the (a) M2 mirror diameter, (b) M3 mirror diameter, and (c) Geometrical collection area of ideal parabola, Hencky, and L'Garde quasi- parabola for R = 50 m and θ = +/-0.5° case.

Although the L'Garde quasi-parabola profile results in smaller M2 and M3 mirror sizes compared to the Hencky surface, they are still larger than those resulting from an ideal parabolic shape. To reduce the required diameter of M2 and M3 further system optimization is required.

Let P1 be a quasi-parabolic reflector profile with base radius of curvature of 50 m at nominal pressure. P1 is decomposed to result in a best fit parabola and the residual W-curve shown in Fig. 7. The W-curve is fit to an 8$^{th}$ order polynomial and a transverse ray analysis performed to identify the individual contribution of aspheric coefficients to the overall aberration induced by P1. Total aberration induced by P1 is equal to the combination of defocus ($w_{020}$), aberration induced by 4$^{th}$ order term of W-curve ($w_{040}$), and aberration induced by the 6$^{th}$ order term of the W-curve ($w_{060}$) as shown in Eq (7),

$$\varepsilon_y = -2 \times f/\# \times \left(2w_{020}\rho^2 + 4w_{040}\rho^4 + 6w_{060}\rho^6\right). \tag{7}$$

The 8$^{th}$ order term is neglected as its contribution to the overall aberration is small compared to the other terms.

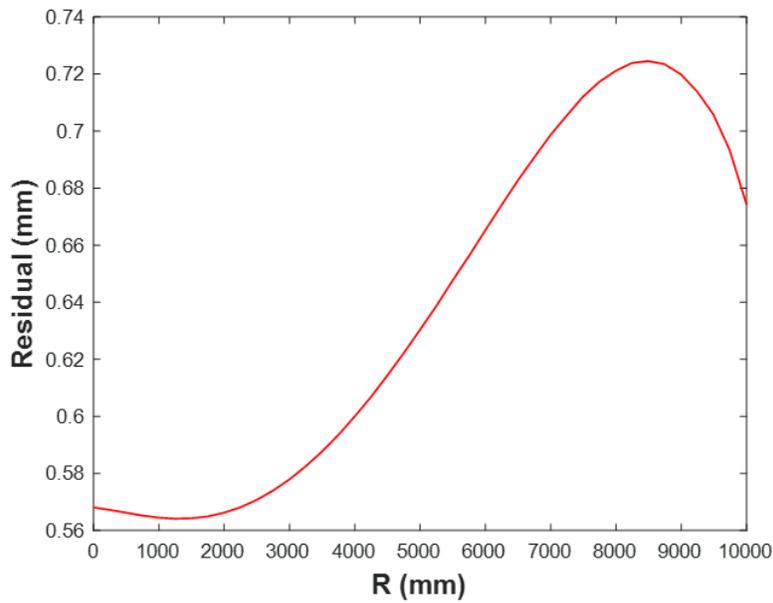

**Fig. 7** W- curve: L'Garde near-parabola data of A1 profile is fit to an ideal parabola with base radius of 50,033.53 mm and the residual error is plotted as a function of radial distance from the A1 optical axis. (Note: This radial plot only shows the half of the "W" shaped W-curve.)

The transverse ray analysis yields values of, $\varepsilon_y = 671.3$ mm, $w_{040} = 10.4$ mm, $w_{060} = 12.9$ mm and $f/\# = 1.6$. Substitution into Eq (7) then yields

$$w_{020} = 164.4 \text{ mm}.$$

Defocus affects the location of mirrors and can be accounted for during the design process. The spherical aberration terms $w_{040}$ and $w_{060}$ affect the size of M2-M3 corrector mirror sizes. Reduction in the contribution of these two terms results in smaller mirrors. A scaling factor is applied to A4 and A6 aspheric terms to demonstrate its effect on M2 and M3 sizes (Table 4).

Table 4 Variation of M2 diameter as a function of scaling factor applied to the L'Garde quasi-parabola A4 and A6 aspheric coefficients of the W-curve.

| Scaling Factor | M2 Diameter (mm) | Peak to valley error w.r.t best fit parabola (mm) |
|---|---|---|
| 1 | 1322 | 13.53 |
| 0.5 | 592 | 6.9 |
| 0.1 | 456 | 1.6 |
| 0.01 | 370 | 0.4 |

Such scaling factors can be realized by adjusting the pressure within the reflector. L'Garde data for quasi-parabolic profiles with $R_1 = 50$ m, $D_1 = 20$ m at $\pm 10\%$, $\pm 20\%$, and $\pm 30\%$ of nominal pressure were analyzed to see if any of these are a close match to the profiles predicted in Table 4. The profile at 20% lower than nominal pressure results in a M2 diameter of 500 mm and is close to the profile calculated using a scaling factor of 0.1. However, 20% lower than nominal pressure is not sufficient to prevent wrinkles in the membranes. Therefore, a conservative scaling factor of 0.15, corresponding to a pressure yielding a wrinkle free surface, is adopted.

## 5 Inflatable Optical Telescope Design Solution Space

The updated definition of the A1 sag profile is used as input to the iterative analytical model and the resultant output is used to develop corresponding ZEMAX models. The next step involves numerical evaluation of the performance of these A1 profiles over the entire FOV. The goal is to efficiently couple the photons collected by the telescope into the focal plane instruments (i.e., the heterodyne receivers) over all the observation bands. The OASIS receiver band wavelength definitions are listed in Table 5.

| **OASIS Receiver Band** | **Wavelength** |
|---|---|
| 1 | 660 - 520 μm |
| 2 | 272 - 136 μm |
| 3 | 120 - 103 μm |
| 4 | 63 μm |

Table 5. OASIS receiver bands

The effective collection area is defined as the product of geometrical photon collection area and the area-averaged Strehl Intensity Ratio (SIR) and is a function of the coupling efficiency from the telescope optics to the beam expected by the heterodyne receivers (F/16 for OASIS). The SIR can be approximated as[14]

$$SIR = e^{-\left(2\pi\frac{\sigma}{\lambda}\right)^2}, \qquad (8)$$

where $\sigma$ is RMS (Root Mean Square) wavefront error as designed. Since the design accommodates FOV of ±0.05°, $\sigma$ averaged over FOV is used for Eq (8). Under the assumption of $\sigma$ quadratically increasing over the FOV,

$$\sigma(r) = \left(\sigma_{off-axis} - \sigma_{on-axis}\right)r^2 + \sigma_{on-axis}, \qquad (9)$$

where $r$ is the normalized field FOV, $\sigma_{on-axis}$ is the RMS wavefront error at FOV = 0 deg, and $\sigma_{off-axis}$ is the RMS wavefront error at FOV = 0.05°. Area-averaged SIR is calculated as

$$SIR(r) = 2\int_0^1 r e^{-\left(2\pi\frac{\left(\sigma_{off-axis}-\sigma_{on-axis}\right)r^2+\sigma_{on-axis}}{\lambda}\right)^2} dr. \qquad (10)$$

The effective photon collection area, EA, of the system is then calculated as

$$EA(r) = 2 \times A_g \int_0^1 r e^{-\left(2\pi \frac{\left(\sigma_{off-axis} - \sigma_{on-axis}\right) r^2 + \sigma_{on-axis}}{\lambda}\right)^2} dr, \tag{11}$$

where $A_g$ is the geometrical photon collection area.

Fig. 8 shows contour plots of the resulting solutions space over all the four observation bands. The effective photon collection area, along with the diameters of M2 and M3, as a function of reflector radius of curvature (R1) and effective pupil diameter (EPD). The effective areas of Bands 1, 2, and 3, are similar to each other, since the wavelength is much longer than the as-design optical path difference of wavefront error. Whereas the Band 4 contour plots suffers from the decreased coupling efficiency due to the shorter wavelength.

As a test of design robustness A1, M2, and M3 are individually perturbed by introducing decenter in X, Y, and Z directions and tilt about X and Y axes with values tabulated in Table 6. The system RMS wavefront error is then estimated by the Root Sum Squares (RSS) rule of the wavefront errors of individual optical components under perturbations. The system RMS wavefront error ($\sigma_{sys}$) with respect to decenter in X, Y, and Z directions, tilt about X and Y axes of individual optical elements is given by

$$\sigma_{sys} = \sqrt{\sigma_{M1}^2 + \sigma_{M2}^2 + \sigma_{M3}^2 + \sigma_{Baseline}^2}, \tag{12}$$

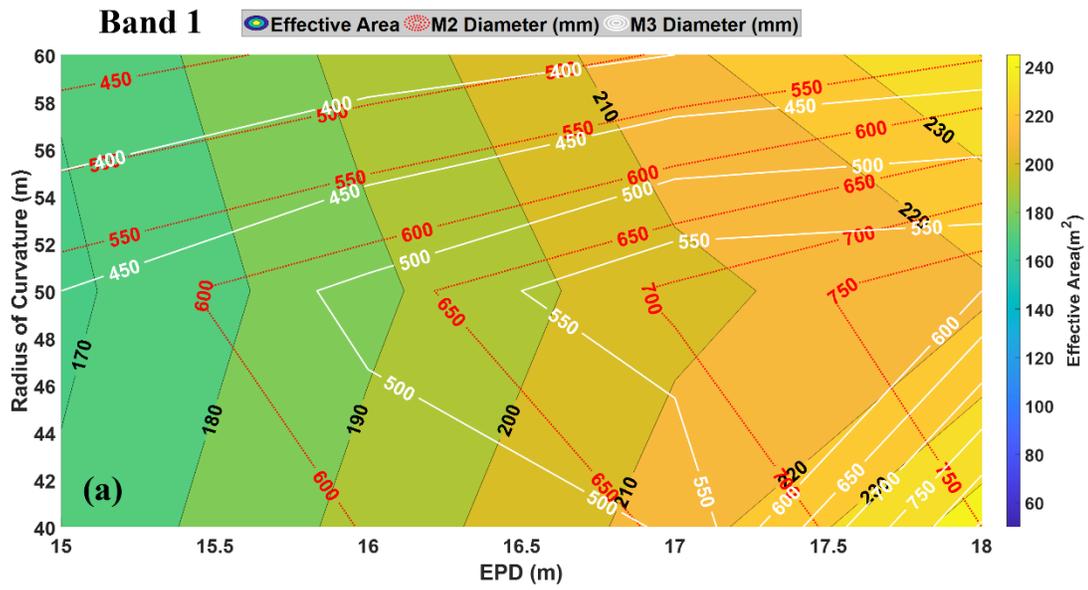
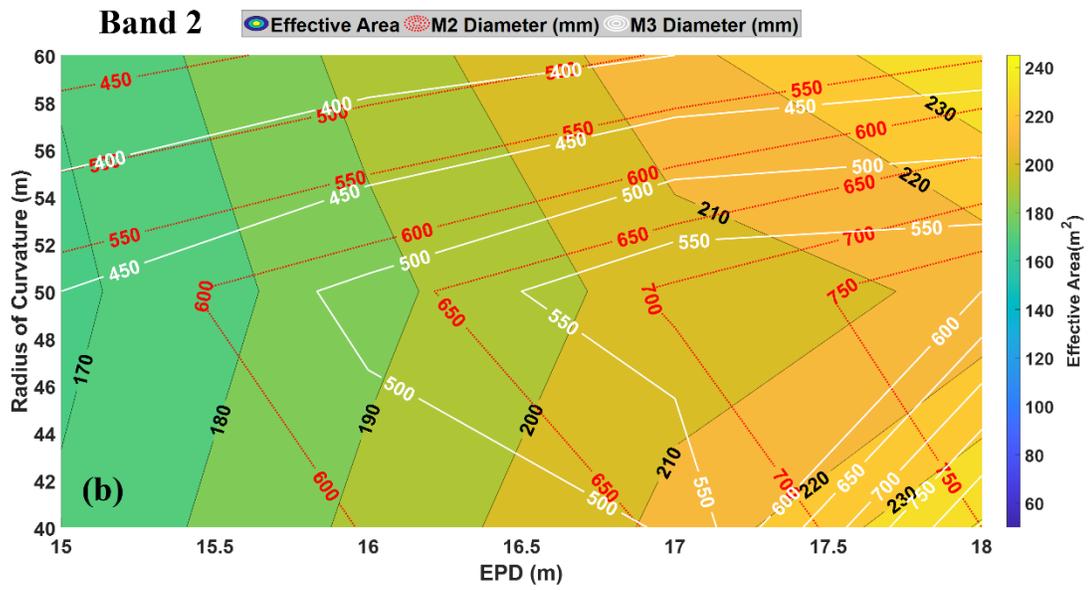

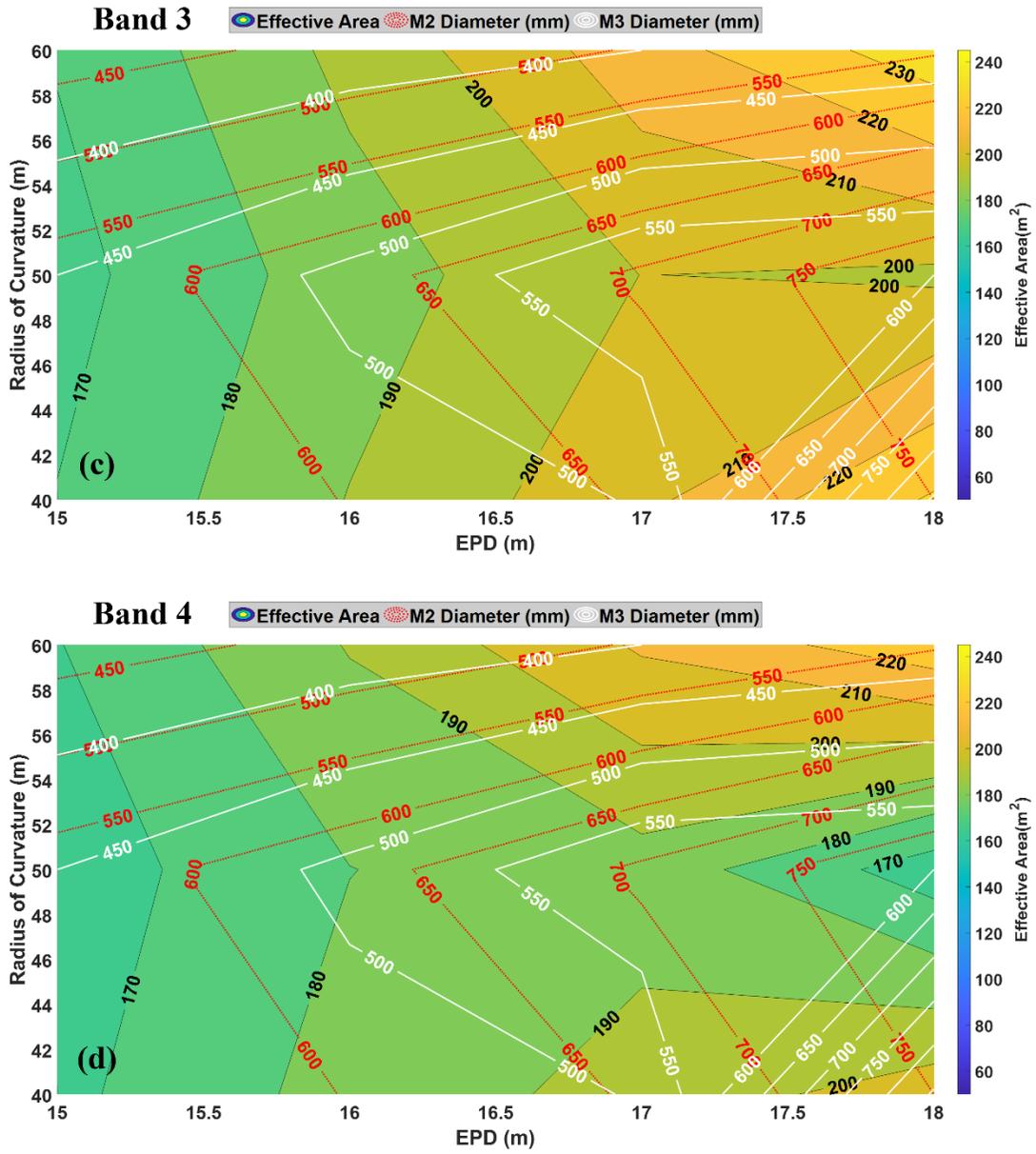

**Fig. 8(a) to (d)** Inflatable optical design solution space contour plots of as designed models over the four OASIS observation bands. Effective photon collection area and diameters of M2 and M3 mirrors are plotted as a function of A1 radius of curvature and EPD.

where the RMS wavefront error of each individual optical elements is

$$\sigma_{element} = \sqrt{\sigma_{Dec\_x}^2 + \sigma_{Dec\_y}^2 + \sigma_{Dec\_z}^2 + \sigma_{Tilt\_x}^2 + \sigma_{Tilt\_y}^2} \ . \tag{13}$$

$\sigma_{sys}$ is substituted in Eq (11) and the results are shown in Fig. 8.

The contour plots show that the effective collecting area decreases due to wave front aberrations and hence the system performance degrades as the wavelength gets shorter. Tolerancing helps in determining the sensitivity of different bands to the overall system RMS wavefront error due to the perturbations of the individual optical elements. Fig. 9 shows that Band 1 and 2 are less susceptible due to their longer wavelengths. After tolerancing, Band 4 shows significant degradation in the effective photon collection area. Depending on the importance of each band (e.g., concept of operations, quality and number of astronomical targets, integration time), the solution space can be recalculated using a weighing factor. The combination of A1 radii of curvatures and entrance pupil diameters which would satisfy the science goals (effective collection area) and is within system architecture constraints (M2-M3 mirror sizes) can be then identified by comparing these weighted solution space contour plots. The systematic optimization flow involved in the parametric design space search resulting in solution space contour plots are presented in Fig. 10.

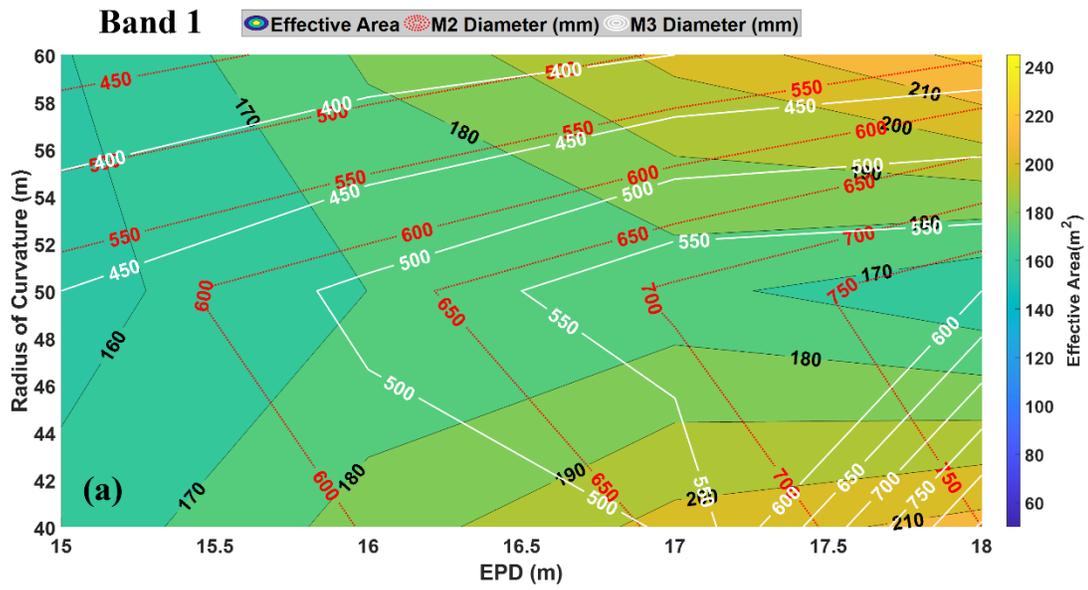

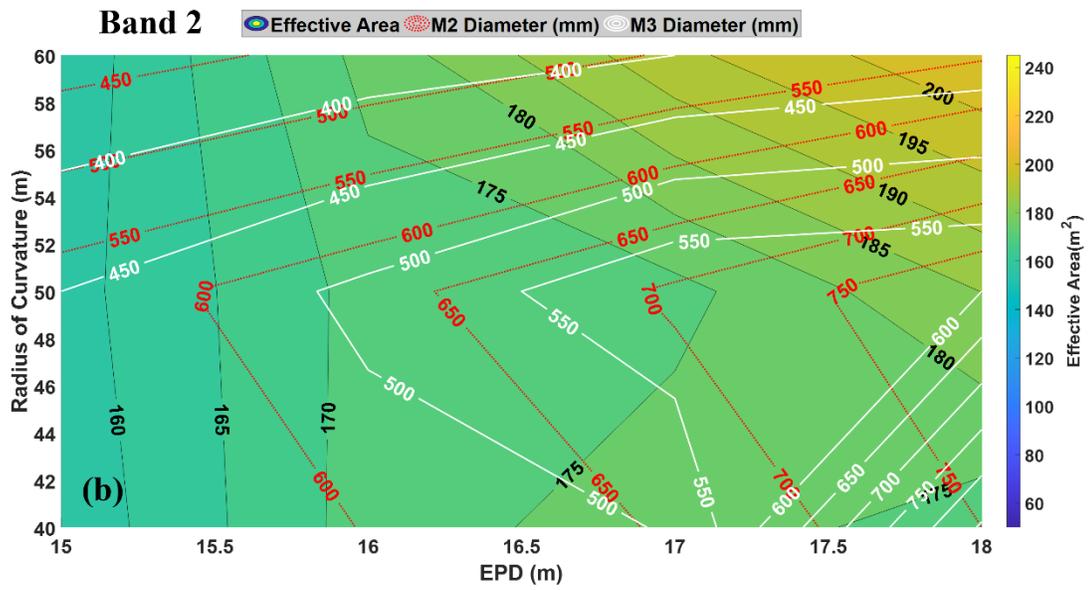

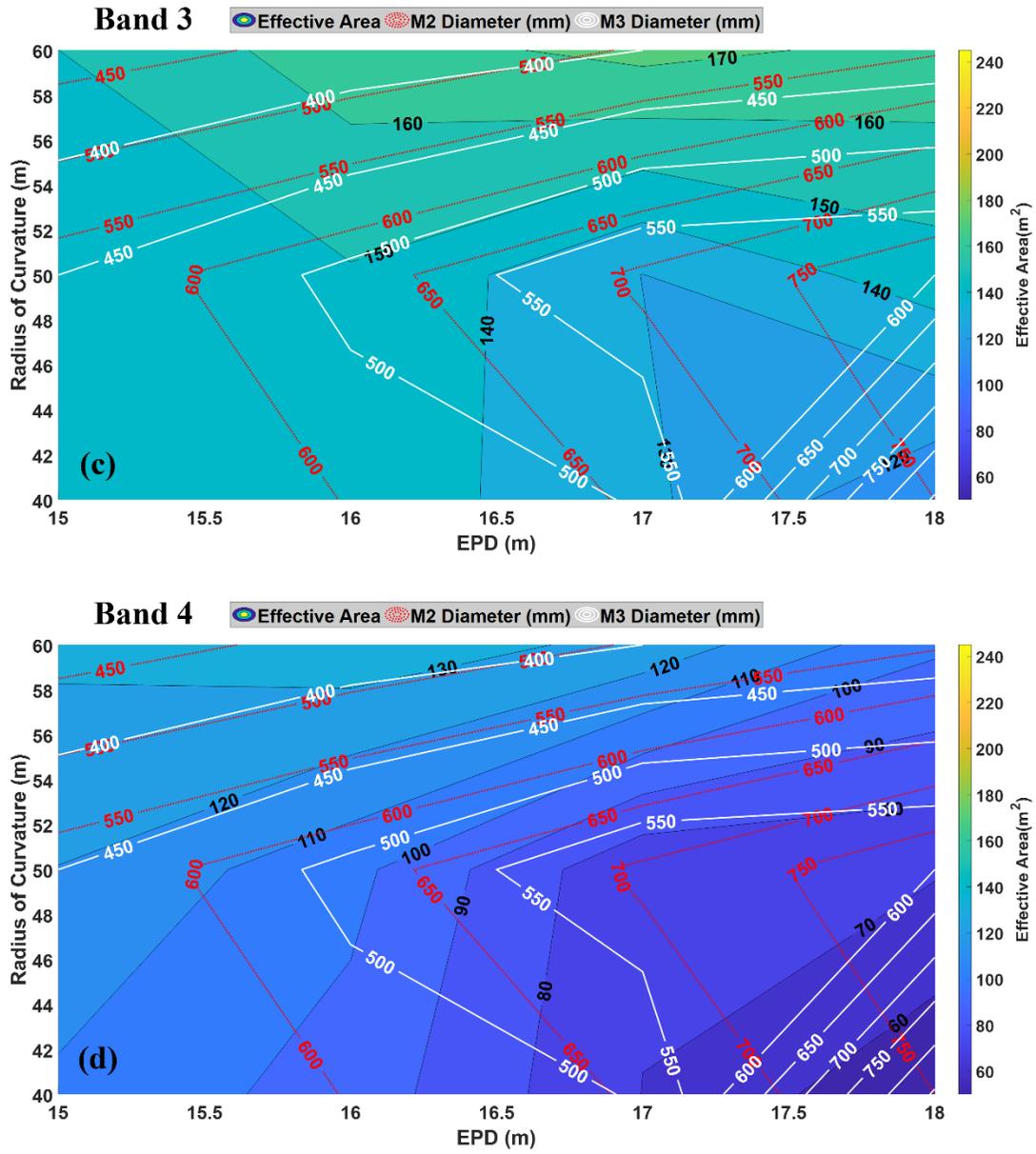

**Fig. 9(a) to (d)** Inflatable optical design solution space contour plots post tolerancing over the four OASIS observation bands. Effective photon collection area using Eq. (11) and diameters of M2 and M3 mirrors are plotted as a function of A1 radius of curvature $R_1$ and EPD

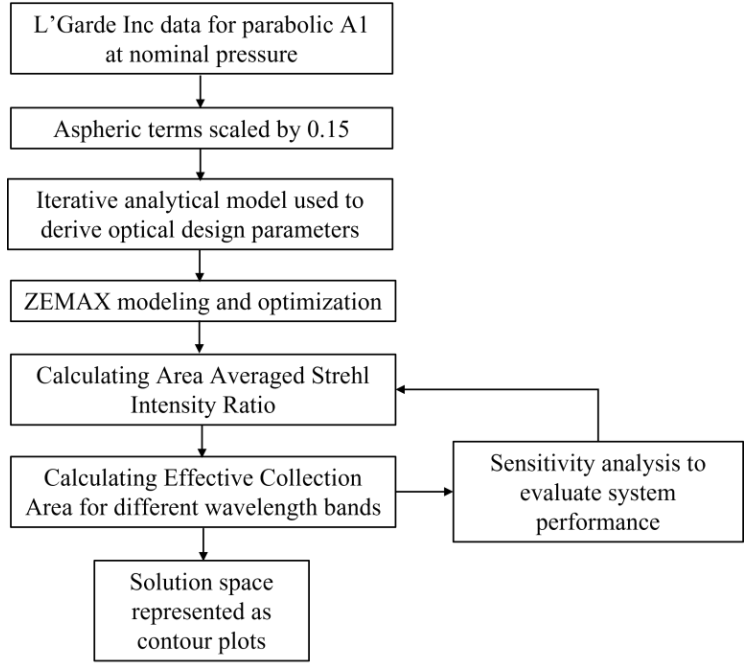

**Fig. 10** OASIS parametric design space search process and inflatable optical design optimization workflow.

**Table 6**. Baseline design parameters and the OASIS optical performance sensitivity analysis parameters.

| Parameter | | Value | |
|---|---|---|---|
| $R_1$ | | [40m 50m 60m] | |
| $D_1$ | | [15 m 16 m 17 m 18 m] | |
| $d_{2-3}$ | | 0.7 m | |
| $d_{IF}$ | | 100 mm | |
| $\Theta$ | | ±0.05° | |
| A1 | Decenter X = 0.5 mm | Tilt X = 0.001° | |
| | Decenter Y = 0.5 mm | Tilt Y = 0.001° | |
| | Decenter Z = 0.5 mm | Tilt Z = 0° | |
| M2 | Decenter X = 0.1 mm | Tilt X = 0.001° | |
| | Decenter Y = 0.1 mm | Tilt Y = 0.001° | |
| | Decenter Z = 0.5 mm | Tilt Z = 0° | |
| M3 | Decenter X = 0.1 mm | Tilt X = 0.001° | |
| | Decenter Y = 0.1 mm | Tilt Y = 0.001° | |
| | Decenter Z = 0.5 mm | Tilt Z = 0° | |

As an example of design optimization utilizing Fig. 9, let us assume that due to the requirement on the maximum mass of inflatant and the rate of change of surface profile with respect to pressure, a radius of curvature of 50 m is selected for A1. The corresponding data from the contour plots (Fig 9a to 9d) for $R_1$ = 50 m are listed in Table 7.

**Table 7**. Effective area and mirror diameters for $R_1$ = 50 m.

| EPD (m) | Effective Area (m²) | | | | Mirror Diameter (mm) | |
| --- | --- | --- | --- | --- | --- | --- |
| | Band 1 | Band 2 | Band 3 | Band 4 | M2 | M3 |
| 15 | 156 | 158 | 147 | 119 | 574 | 450 |
| 16 | 170 | 171 | 149 | 103 | 634 | 510 |
| 17 | 173 | 173 | 129 | 71 | 708 | 590 |
| 18 | 161 | 185 | 145 | 71 | 792 | 600 |

Considering the maximum effective photon collection area for each band and ignoring the constraint on mirror diameter, Band 1 has a peak at 17 m EPD, Band 2 at 18 m, Band 3 at 16 m, and Band 4 at 15 m. From Table 7, an A1 with EPD of 16 m for $R_1$ = 50 m is the best choice as it still delivers close to maximum possible EAs at Band 1, 2, and 4. But if achieving the smallest possible mirror diameter has higher priority than the maximum possible EA, EPD of 15 m for $R_1$ = 50 m turns out to be the best option. Thus, plots of the type shown in Fig. 9 help in efficiently navigating the multi-variable dependencies of the design space and in identifying the most suitable A1 profile.

## 6  FOV Scanner Design

A Field of View (FOV) of ±0.05° (3 arc-minutes) is part of the OASIS design specifications in Table 1. This is achieved by scanning the intermediate image with a tip-tilt mirror. The stop is placed at A1 in the OASIS design. A field lens made of HRFZ Silicon is placed at the intermediate focal plane to minimize the size of the scanning mirror and to avoid vignetting during scanning.

Absorption loss of HRFZ Si could be a cause of concern. But since the central thickness of the field lens is sufficiently thin, 2 mm, the maximum absorption loss is at most 15 %, which occurs in Band 4 (Fig. 11). Reflection loss is mitigated by using the right combination and thickness of AR coating material such as Parylene[15].

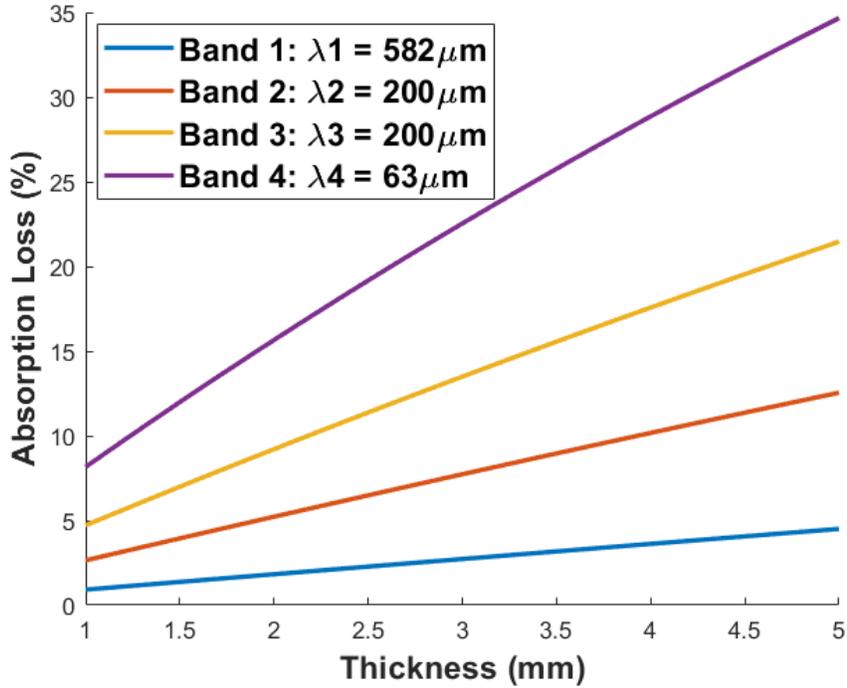

**Fig. 11** Absorption loss of HEFZ Si as a function of central thickness of lens.

The optical layout of the scanning mechanism is shown in Fig. 12a. The tip-tilt mirror M4 scans the intermediate image and then it is re-imaged by an ellipsoidal mirror M5 to achieve the F/16 system as per the OASIS design specification in Table 1.

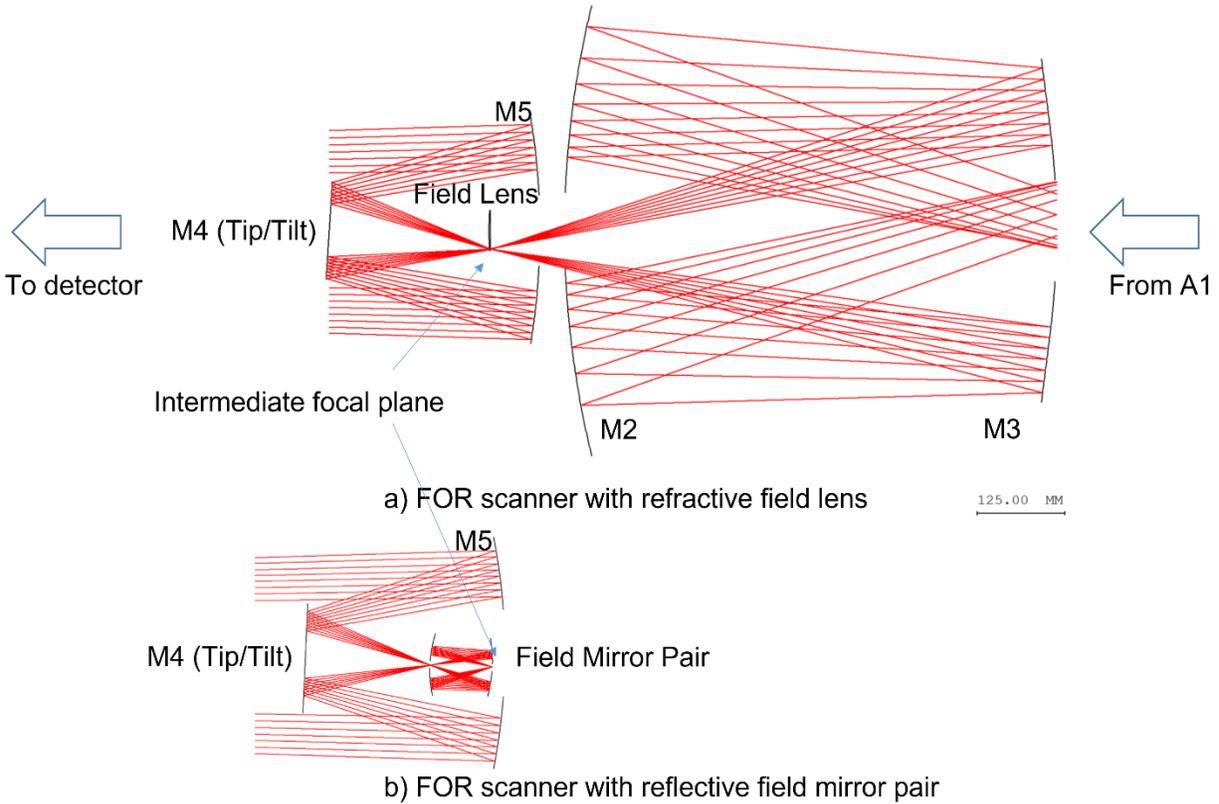

**Fig. 12** OASIS FOV scanner layout with (a) HRFZ Si field lens, and (b) mirror pair (RFL1 and RFL2) replacing the field lens.

If the absorption loss of 15% at Band 4 needs to be eliminated, the field lens can be replaced with a pair of mirrors as shown in Fig. 12b. The mirror pair is rigidly connected to the tip/tilt M4 and shares the same axis about which they are rotated. In addition to the reduced loss, this configuration can also support a wider FOV of ±0.1°.

## 7  MidEx Class 14 m diameter OASIS Optical Design

*7.1 Optical Design*

Driven by the heritage of large inflatable mirror[16] and system architecture constraints[9], the diameter of A1 was selected to be 14 m for the NASA MidEx proposal[2]. The methodologies detailed in previous sections are used to redesign and optimize the corrector module, FOV scanner, and calculate the effective collection area for all the bands.

L'Garde Inc provided new data for an A1 diameter of 14 m and Radius of Curvature (RoC) = [40 m, 50 m, and 60 m]. The data was fit to an $8^{th}$ order polynomial and the aspheric coefficients scaled by 0.15 to produce the new A1 surface profiles. The iterative analytical model was used to design the corrector module (M2-M3). Table 8 shows the preliminary modeling results by ray-tracing software (ZEMAX) for all the three radii of curvatures. The RoC of 40 m provides the largest geometric photon collection area with the smallest mirror dimensions.

**Table 8** ZEMAX modeling results for different A1 radii of curvatures.

| Radius of Curvature (m) | M2 Radius (mm) | M2 Hole Radius (mm) | M3 Radius (mm) | M3 Hole Radius (mm) | M2-M3 Distance (mm) | Geometric Collection Area (m$^2$) |
|---|---|---|---|---|---|---|
| 40 | 240 | 50 | 216 | 50 | 700 | 142.77 |
| 50 | 265 | 50 | 195 | 65 | 700 | 126.93 |
| 60 | 241 | 50 | 170 | 75 | 700 | 100.62 |

The FOV scanning system shown in Fig. 12 was redesigned to be compatible with a commercially available hexapod (M4) H-811 from PI[17]. To prevent vignetting due to the hexapod and considering mounting constraints, M4 is mounted 45° off-axis. M5-M6 mirror pairs are used for further correction and to achieve f/16 system. Three flat mirrors, M7, M8, and M9 are used for folding the beam inside the packaging volume. Fig. 13 and 14 shows the MidEx Class OASIS optical design using 14 m diameter inflatable primary antenna and the spot diagrams showing diffraction-limited optical performance at $\lambda = 111$ μm (Band 3).

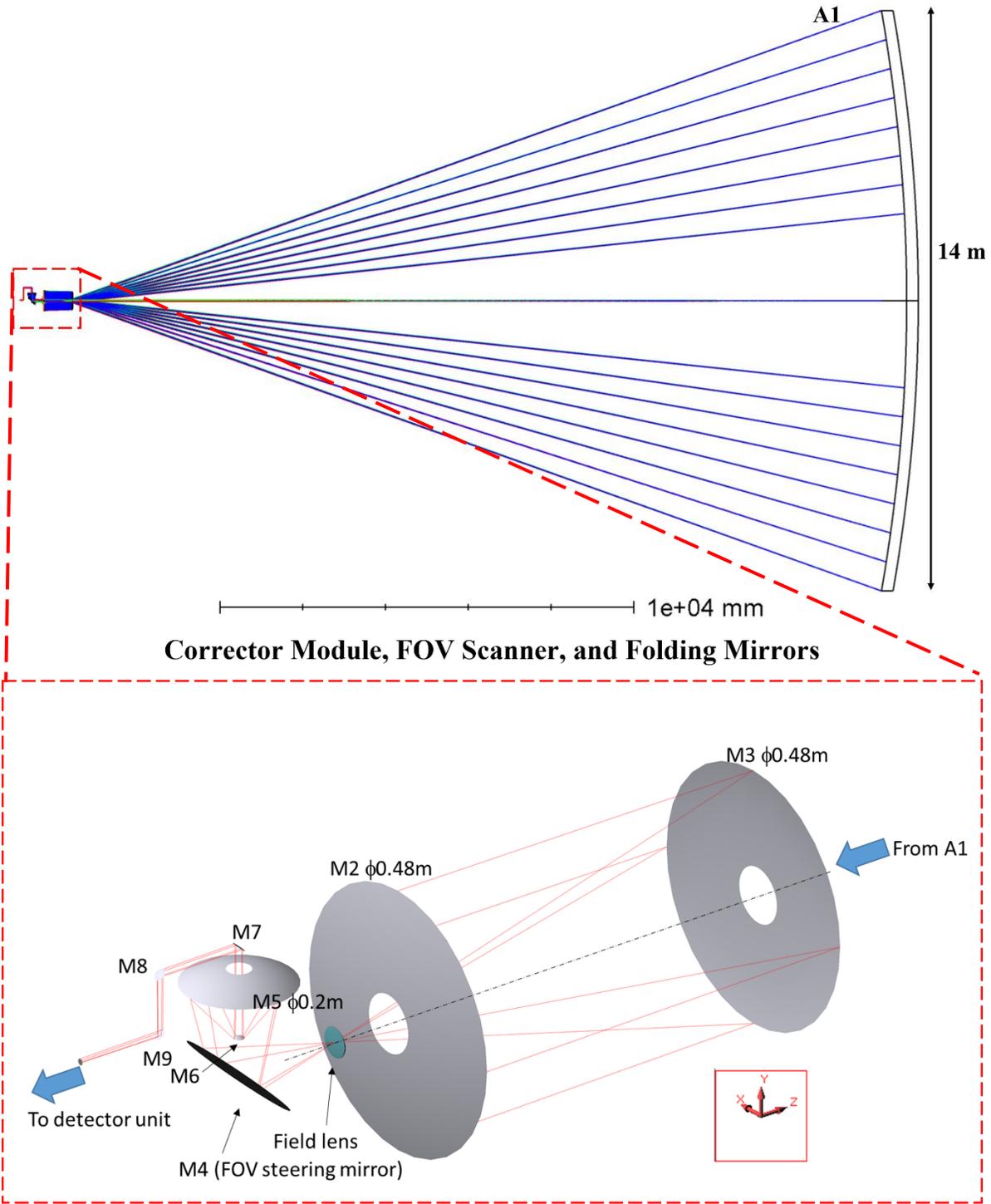

**Fig. 13** Ray-trace model showing the MidEx Class OASIS optical design including the corrector module, field lens, FOV scanner, and folding mirrors for 14 m diameter primary antenna A1.

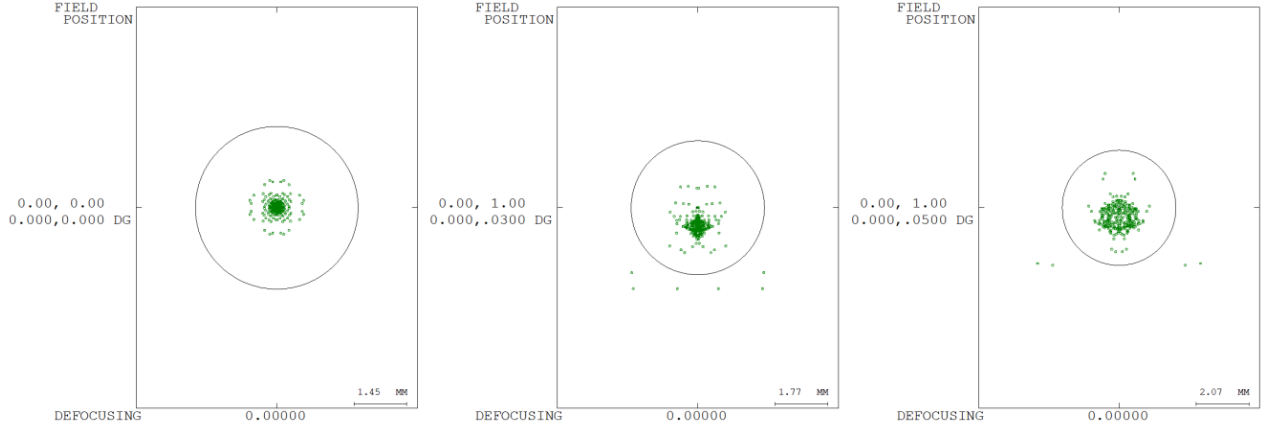

**Fig. 14** MidEx Class 14 m diameter OASIS spot diagram showing diffraction limited performance at 0°, 0.03°, and 0.05° FOV for λ = 111 μm (Band 3).

A sensitivity analysis was performed, and effective collecting areas calculated as described in Section 5. Table 9 provides the effective photon collecting area derived using Eq (11), along with the aperture efficiency, which is the ratio between the Effective Collection Area (*EA*) and the Geometrical Membrane Area ($A_g$).

**Table 9** Effective photon collection area of the MidEx Class 14 m diameter OASIS.

| Band | λ (μm) | Effective Collection Area (m²) | Membrane Area (m²) | Aperture Efficiency |
|---|---|---|---|---|
| 1 | 582 | 142 | **154** | **0.922** |
| 2 | 200 | 140 | **154** | **0.909** |
| 3 | 111 | 130 | **154** | **0.844** |
| 4 | 63 | 108 | **154** | **0.701** |

*7.2 OASIS End-to-End Optics Loss Budget*

OASIS employs terahertz heterodyne receivers which require dichroics, beam splitters, and reimaging optics to couple the beam from the intermediate focal point to the focal plane mixers. These components reside in the Receiver Module (Fig. 2). A block diagram of the OASIS back-end receiver architecture is shown in Fig. 15.

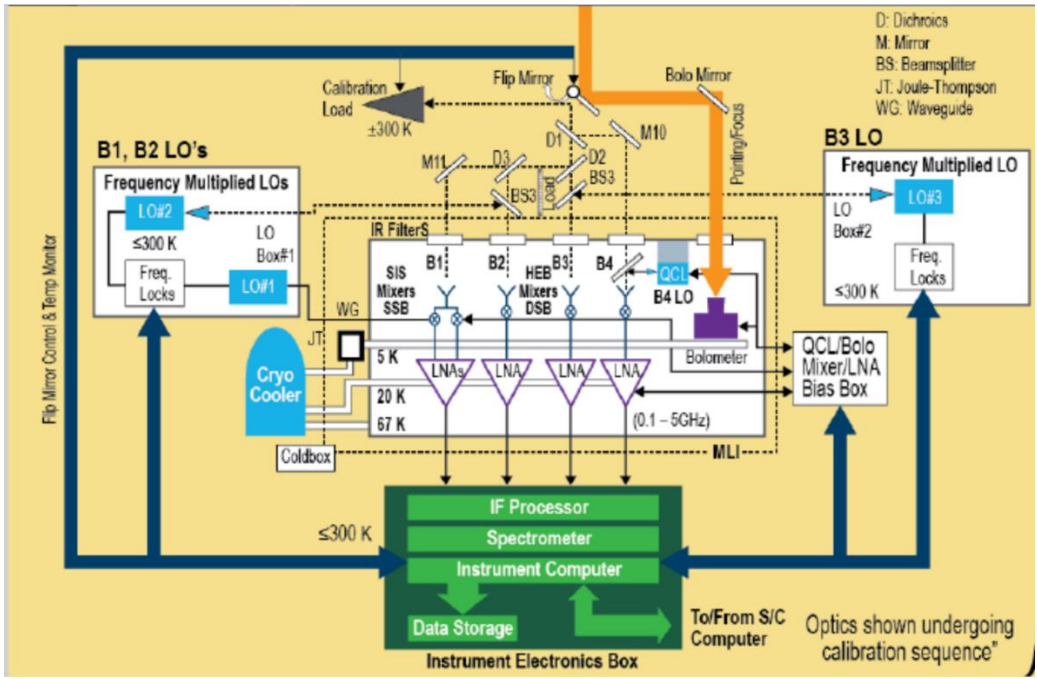

**Fig. 15** Block diagram of OASIS receiver architecture.

Estimates for the absorption and scattering losses associated with A1, the corrector module, and receiver optics are listed in Table 10. These values are included in determining the end-to-end transfer efficiency ($\varepsilon_T$) and *EA* of the system. The results are summarized in Table 10.

The Aperture Efficiency in Table 8 was used to estimate the effective photon collection area while considering all the optical components depicted in Fig. 13 and 15. Our analysis indicates a 14 m diameter OASIS meets mission science requirements with margin.

Table 10. OASIS optics loss budget.

| Module | Element | Type | Material | Loss | Band 1 | Band 2 | Band 3 | Band 4 |
|---|---|---|---|---|---|---|---|---|
| A1 | C1 | Canopy | Transparent Kapton | 0.02 | 0.02 | 0.02 | 0.02 | 0.02 |
| A1 | A1 | Reflector | Aluminized Kapton | 0.01 | 0.01 | 0.01 | 0.01 | 0.01 |
| Corrector | M2 | Asphere | Aluminum | 0.01 | 0.01 | 0.01 | 0.01 | 0.01 |
| Corrector | M3 | Asphere | Aluminum | 0.01 | 0.01 | 0.01 | 0.01 | 0.01 |
| Corrector | FL | Field Lens | Si | 0.02 | 0.02 | 0.02 | 0.02 | 0.02 |
| Corrector | M4 | Flat: FOV scanner | Aluminum | 0.01 | 0.01 | 0.01 | 0.01 | 0.01 |
| Corrector | M5 | Asphere | Aluminum | 0.01 | 0.01 | 0.01 | 0.01 | 0.01 |
| Corrector | M6 | Asphere | Aluminum | 0.01 | 0.01 | 0.01 | 0.01 | 0.01 |
| Corrector | M7 | Flat | Aluminum | 0.01 | 0.01 | 0.01 | 0.01 | 0.01 |
| Corrector | M8 | Flat | Aluminum | 0.01 | 0.01 | 0.01 | 0.01 | 0.01 |
| Corrector | M9 | Flat | Aluminum | 0.01 | 0.01 | 0.01 | 0.01 | 0.01 |
| Receiver | D1 | Frequency Diplexer | Mesh | 0.05 | 0.05 | 0.05 | 0.05 | 0.05 |
| Receiver | D2 | Frequency Diplexer | Mesh | 0.05 | 0.05 | 0.05 | 0.05 | - |
| Receiver | D3 | Frequency Diplexer | Mesh | 0.05 | 0.05 | 0.05 | - | - |
| Receiver | B2 | Beam Splitter | Mylar | 0.01 | - | 0.01 | - | - |
| Receiver | B3 | Beam Splitter | Mylar | 0.01 | - | - | 0.01 | |
| Receiver | B4 | Beam Splitter | Mylar | 0.01 | - | - | - | 0.01 |
| Receiver | F1 | IR Filter | Mesh | 0.02 | 0.02 | - | - | - |
| Receiver | F2 | IR Filter | Mesh | 0.02 | - | 0.02 | - | - |
| Receiver | F3 | IR Filter | Mesh | 0.02 | - | - | 0.02 | - |
| Receiver | F4 | IR Filter | Mesh | 0.02 | - | - | - | 0.02 |
| Receiver | M10 | Flat | Aluminum | 0.01 | - | - | - | 0.01 |
| Receiver | M11 | Elliptical | Aluminum | 0.02 | 0.02 | - | - | - |
| | | Transmission efficiency | | | 0.68 | 0.69 | 0.74 | 0.78 |
| | | Aperture Efficiency | | | 0.922 | 0.909 | 0.844 | 0.701 |
| | | Transfer Efficiency | | | **0.63** | **0.63** | **0.62** | **0.54** |
| | | Membrane Area | | | 154 | 154 | 154 | 154 |
| | | Effective Collection Area | | | 96.6 | 96.6 | 96.0 | 84.2 |
| | | Requirement (10x Herschel) | | | 56 | 56 | 56 | 56 |
| | | Margin | | | **42%** | **42%** | **42%** | **34%** |

*7.3 Resolution of A1 Pressure Control Unit*

Since the A1 surface profile, and location of intermediate focal plane of A1-M2-M3 is a function of A1 pressure, the resolution of the pressure control unit becomes a critical parameter to match telescope beam waist at the location of heterodyne receivers. Data from L'Garde Inc for D = 14 m, R = 40 m at nominal, 10 % higher, and 10% lower than nominal pressure is analyzed to determine the rate of change of base radius of curvature with respect to change of pressure. For this configuration, numerical simulation shows $dr/dp = 1.72$ mm/mPa. Considering a resolution of 0.1% of nominal pressure,

$$dp = 0.1\% \text{ of } P = 0.1\% \text{ of } 4.37 \text{ Pa} \approx 5 \text{ mPa},$$
$$\therefore dr = 8.6 \text{ mm}.$$

The current design can accommodate $dr$ (displacement of A1 along Z-axis) of 0.5 mm (Table 6). A maximum change of $dr = 2$ mm can be compensated for by using the M4 hexapod. Thus, a pressure control unit with a resolution of 1 mPa is required to maintain optimum system performance. If actuators with a travel $\pm 1.65$ mm are incorporated into the boom design, then a resolution of 5 mPa will suffice.

## 8 Discussion

Starting with analytically or numerically derived surface profiles, this work has shown it is possible to realize high performance telescopes with large, inflated apertures. The next challenge from the optical design viewpoint is the compensation for thermal deformation (or unforeseen aberrations) in A1 by implementing adaptive optics (AO). AO is not required for OASIS but may prove necessary for future telescopes with inflated primaries operating at shorter wavelengths. AO is widely used on ground based telescopes to compensate for rapid variations in the refractive index of the Earth's atmosphere due to turbulence. The changes in surface figure expected in space based inflated optics due to thermal effects will occur over minutes or hours, reducing the complexity of and demand on the AO system. In addition, the shape of A1 can be controlled by precisely modulating the pressure. This can be used in conjunction with the AO to correct for thermal variations in A1.

**Conclusion**

The Orbiting Astronomical Satellite for Investigating Stellar Systems (OASIS) is an inflatable terahertz space telescope with a 14 m diameter primary reflector realized with a pressurized, metalized polymer membrane. We established a systematic optical design process for optimizing the performance of space telescopes employing inflatable primary reflectors of various sizes. The formulation enables a comparison of inflatable primary reflector geometries, such as Hencky reflectors formed from a monolithic flat gore, as well as quasi-parabolic reflectors formed from segmented and sealed gores. The comprehensive design process identified a baseline design for OASIS satisfying all major requirements, such as effective photon collection area and size of corrector optics, as well as tolerancing requirements. Thus, this work provides a roadmap for addressing the unique challenges associated with realizing the great potential of large inflatable optics for space applications.

*References*

1. W. Martin, J. Baross, D. Kelley, and M. J. Russell, "Hydrothermal vents and the origin of life," Nat. Rev. Microbiol., vol. 6, no. 11, pp. 805–814, 2008, doi: 10.1038/nrmicro1991.
2. 2021 Medium Explorer (MIDEX) – Solicitation No. NNH21ZDA018O.

3. Christopher Walker, et al., "Orbiting Astronomical Satellite for Investigating Stellar Systems (OASIS) Following the Water Trail from the Interstellar Medium to Oceans," SPIE Paper #11820-26, Proc SPIE, Optics+Photonics, San Diego, California, 1-5 August 2021.

4. G. L. Pilbratt *et al.*, "Herschel Space Observatory," *Astron. Astrophys.*, vol. 518, no. 7–8, 2010, doi: 10.1051/0004-6361/201014759.

5. Hencky, H., "Uber den Spannungszustand in kreisrunden Platten," Z. Math. Phys. 63, 311-317 (1915).

6. W.B. Fichter, "Some Solutions for the Large Deflections of Uniformly Loaded Circular Membranes," NASA Technical Paper 3658, July 1997

7. Arthur L. Palisoc, et al, "Analytical and finite element analysis tool for nonlinear membrane antenna modeling for astronomical applications", SPIE Paper #11820-32, Proc SPIE, Optics+Photonics, San Diego, California, 1-5 August 2021.

8. James H. Burge, S. Benjamin, M. Dubin, A. Manuel, M. Novak, C. J. Oh, M. Valente, C. Zhao, J. A. Booth, J. M. Good, Gary J. Hill, H. Lee, P. J. MacQueen, M. Rafal, R. Savage, M. P. Smith, B. Vattiat, "Development of a wide-field spherical aberration corrector for the Hobby Eberly Telescope," Proc. SPIE 7733, Ground-based and Airborne Telescopes III, 77331J (5 August 2010); https://doi.org/10.1117/12.857835

9. Jonathan W. Arenberg, Michaela N. Villareal, Jud Yamane, et al, "OASIS architecture: key features," Proc. SPIE 11820, Astronomical Optics: Design, Manufacture, and Test of Space and Ground Systems III, 118200S (24 August 2021); https://doi.org/10.1117/12.2594681

10. J. R. G. Silva, B. Mirzaei, W. Laauwen, N. More, A. Young, C. Kulesa, C. Walker, A. Khalatpour, Q. Hu, C. Groppi, J. R. Gao, "4×2 HEB receiver at 4.7 THz for GUSTO," Proc. SPIE 10708, Millimeter, Submillimeter, and Far-Infrared Detectors and Instrumentation for Astronomy IX, 107080Z (9 July 2018); https://doi.org/10.1117/12.2313410

11. A. B. Meinel and M. P. Meinel, "Inflatable membrane mirrors for optical passband imagery," *Opt. Eng.*, vol. 39, pp. 541–550, 2000, doi: 10.1117/1.602393.

12. Offner A., "A Null Corrector for Paraboloidal Mirrors," Appl. Opt. 2, 153-155 (1963).

13. Greivenkamp, John E. Field Guide to Geometrical Optics. Bellingham: SPIE, 2004. Web.

14. Virendra N. Mahajan, "Strehl ratio for primary aberrations in terms of their aberration variance," J. Opt. Soc. Am. 73, 860-861 (1983).

15. A. J. Gatesman, J. Waldman, M. Ji, C. Musante and S. Yagvesson, "An anti-reflection coating for silicon optics at terahertz frequencies," in *IEEE Microwave and Guided Wave Letters*, vol. 10, no. 7, pp. 264-266, July 2000, doi: 10.1109/75.856983.

16. R.E. Freeland, G.D. Bilyeu, G.R. Veal, M.D. Steiner, D.E. Carson, "Large inflatable deployable antenna flight experiment results", Acta Astronautica, vol. 41, Issue 4-10, pp 267-277 (1997).

17. https://www.pi-usa.us/en/products/6-axis-hexapods-parallel-positioners/h-811i2-6-axis-miniature-hexapod-700886/

**Siddharth Sirsi** is a PhD candidate in Electrical Engineering. He completed his MS in Optical Sciences at the University of Arizona in 2021. His main areas of research are superconducting transistors, terahertz astronomical instrumentation design, and optical design and metrology of inflatable telescopes. He is part of the Steward Observatory Radio Astronomy Lab and Large Optics and Fabrication and Testing Lab at UofA. He completed his MS in ECE from Arizona State University in 2014. (ssirsi@email.arizona.edu)

**Yuzuru Takashima** is an associate professor at James C. Wyant College of Optical Sciences of University of Arizona. His research focus is MEMS-based lidar for automotives and AR displays for Metaverse, as well as optical design in general including space optics. Prior to joining to the University of Arizona, he was a research staff at Stanford University where he conducted research and development of high density holographic data storage systems and nano-photonic electron beam generators. He was employed as an optical engineer at Toshiba Corporation in Japan and developed ultra-precision manufacturing process for optical components. He is a senior member of OPTICA and SPIE, serving as a co-chair of a conference, Industrial Optical Systems and Devices (iODS). He received B.S. in Physics from Kyoto University in Japan and M.S. and Ph.D. in Electrical Engineering from Stanford University. (ytakashima@optics.arizona.edu)